\documentclass[11pt,3p,review,authoryear]{elsarticle}

\usepackage[utf8]{inputenc}
\usepackage[english]{babel}
\usepackage{fancyhdr}
\usepackage{lmodern}
\usepackage{lscape}
\usepackage{color}
\usepackage[table]{xcolor}
\usepackage{epsfig}
\usepackage{epic}
\usepackage{eepic}
\usepackage{multirow}
\usepackage{indentfirst}
\usepackage{xcolor}
\usepackage{algorithm}
\usepackage{algpseudocode}
\usepackage{amsfonts}
\usepackage{amssymb}
\usepackage{amsmath}
\usepackage{amsthm}
\usepackage{threeparttable}
\usepackage{amscd}

\usepackage{here}
\usepackage{graphicx}
\usepackage{lscape}
\usepackage{mathtools}
\usepackage{breakcites}
\usepackage{microtype}
\usepackage{tabularx}
\usepackage{subcaption}
\usepackage{caption}
\usepackage{longtable}
\usepackage{array}
\usepackage{booktabs}
\usepackage{float}        

\usepackage[colorlinks,allcolors=blue,linktocpage]{hyperref}
\usepackage{thmtools}

\usepackage{booktabs}
\usepackage{longtable}
\usepackage{graphicx} 

\journal{Journal of Econometrics and Statistics}
\biboptions{longnamesfirst}

\begin{document}

\begin{frontmatter}

\title{Measuring General Associations in Time Series: An Adaptation and Empirical Evaluation of the CODEC Coefficient in Determining Autoregressive Dynamics}

  \author[mymainaddress]{Juan Pablo Montaño}

  \author[mymainaddress]{Mario E. Arrieta-Prieto \corref{mycorrespondingauthor}}
 \cortext[mycorrespondingauthor]{Corresponding author}
 \ead{mearrietap@unal.edu.co}

 \address[mymainaddress]{Department of Statistics, Universidad Nacional de Colombia, Bogotá, Colombia, 110131}

\begin{abstract}
Identifying the number of lags to include in an autoregressive model remains an open research problem due to the computational burden of treating it as a hyperparameter, especially in complex models. This study explores model-agnostic association measures, including Pearson, Spearman, and an adaptation of the recently proposed 
 conditional dependence coefficient (CODEC), for guiding lag selection in time series. We adapt and implement the CODEC-based Feature Ordering by Conditional Independence (CODEC-FOCI) algorithm and evaluate its performance through extensive simulations across linear, nonlinear, stationary, nonstationary, seasonal, and heteroskedastic processes. Results show that CODEC outperforms classical correlation-based measures in nonlinear and nonstationary settings, especially for large sample sizes. In contrast, Pearson performs better in purely linear models. Applications to benchmark datasets confirm that the CODEC approach identifies lag structures consistent with those reported in the literature. These findings highlight CODEC’s potential as a practical, model-free tool for exploratory lag identification in time series analysis.
\end{abstract}


\end{frontmatter}

\section{Introduction}

Time series modeling, both for explanatory and predictive purposes, requires constructing a model that relates observations at a given time to previous values. The number of lags, $p$, is considered a structural parameter of time series models and determines how many past observations are needed to explain the observed behavior at a given instant.

Since the introduction of linear ARIMA-type models, mechanisms for identifying $p$ have been proposed using partial autocorrelation functions (PACFs) and information criteria (e.g., AIC or BIC). However, the former approach primarily is motivated by linear relationships, which may limit their applicability, especially in cases where nonlinear models are more appropriate, such as ARIMA-GARCH, TAR, SETAR, STAR, and, more recently, LSTM neural network models. Conditional measures of association, such as Kendall’s tau \citep{kendall1938new, brophy1986algorithm} and Spearman’s rho \citep{zar2014spearman}, provide robust, rank-based tools to quantify monotonic relationships between variables while also being less sensitive to outliers and specifications of the type of non-linearity present in the data. In time series analysis, these measures could be used to explore dependencies between lagged variables and current outcomes of a given variable in a computationally efficient way \citep{kim2015ppcor, artner2022shape}. This is particularly useful in contexts where linear correlation may fail to capture the true dependence patterns, although Kendall's tau and Spearman's rho are limited to detect monotonic relationships.

The use of information criteria, on the other hand, requires the parametric specification of the underlying model, providing results that can be affected by model misspecification. In general, $p$ could be treated as a hyperparameter that is calibrated through validation or cross-validation approaches, but this search process is often computationally expensive because for each candidate value of $p$ the dataset needs to be rearranged to construct the framework of a regression setting in which the explanatory variables (e.g., the $p$ lags of the variables) conform the design matrix used as input to explain the observed outcome at the current time instant. This reconfiguration of a regression setting for each candidate value for $p$ requires substantial computation time, which adds up to the time required to train complex models. In this situation, one might be tempted to consider a very large value of $p$ as a sensible strategy to avoid its calibration, perhaps motivated by the invertibility property of linear ARMA-like processes \citep[p. 25]{wei2006time}. However, this can induce more problems into the model, such as: (i) curse of dimensionality and reduction of the effective sample size for training, (ii) multicollinearity of the features, (iii) overfitting and poor ex-post predictive ability; and (iv) higher computational burden, due to the size of the feature space. As pointed out by \citet{bengio1994learning} and \citet[p. 274]{box2015time}, all gradient-based and statistical learning algorithms have difficulties during training if the number of lags included is unnecessarily large. 

While previous studies have explored the effectiveness of PACFs in identifying $p$ and recommended their use even for fitting nonlinear models \citep[p. 245]{dixon2020machine}, other proposals have aimed to capture more general relationships between variables to determine 
$p$. Among these, the distance covariance function has been proposed as a potential alternative, which is based on the computation of partial characteristic functions for the underlying process \citep{edelmann2019updated}. It is computationally intensive as it requires the estimation of the conditional characteristic functions of the process and the numerical approximation of a double integral to compute each one of its coefficients.

More recently, \citet{chatterjee2021new} proposed a new coefficient to measure association between variables, with a particular emphasis on nonlinear non-monotonical relationships. The Xi coefficient ($\xi$ when symbolized with a greek letter) is estimated via a transformation of the original data into ranks.

Let $\left\{ (X_i,Y_i)\right\}_{i=1}^n$ be a random sample of a bivariate population in which it is assumed that there are no ties for the values of $X$ (the article provides a generalization in case there are ties). After rearranging the sample as $(X_{(1)},Y_{(1)}),(X_{(2)},Y_{(2)}),\ldots,(X_{(n)},Y_{(n)})$; so that $X_{(1)}<X_{(2)}<\ldots<X_{(n)}$, let $r_i$ be the rank that occupies $Y_{(i)}$ within the values of $Y$. \citet{chatterjee2021new} proposed the sample-based measure of association

\begin{equation*}
    \hat \xi_n (X,Y)= 1-\frac{3 \sum_{i=1}^n{\left|{r_{i+1}-r_{i}}\right|}}{n^2-1}.
\end{equation*}

The author showed that this measure of association converges in probability to a parameter, $\xi(X,Y)$, that lies between 0 and 1; being 0 when $X$ and $Y$ are independent, and 1 when the two variables follow a deterministic equation of the form $Y=f(X)$, for some measurable function $f$. It is important to highlight that, unlike other coefficients of association such as Pearson's correlation coefficient, Kendall's tau, Spearmann's rank correlation coefficient, and even the distance covariance-based coefficient; this recently proposed coefficient is not symmetric with respect to its arguments, i.e., $\xi(X,Y) \neq \xi(Y,X)$. That is particularly relevant because one of the main advantages of this new coefficient is that it quantifies the existence of any sort of functional relationship, being specially powerful to detect smooth and non-monotonic ones between $X$ and $Y$, so the relationship in the other sense (the influence of $Y$ on $X$) might not even be able to be represented by a function.

\citet{azadkia2021simple} extended the use of the $\xi$ coefficient to measure conditional dependencies. The idea is to measure the functional dependency of $Y$ on $X$ conditioned on a vector of other variables, $\mathbf{Z}$, to isolate the marginal contribution of $X$. This conditional coefficient is relabeled as CODEC (Conditional dependence coefficient). The authors even proposed a methodology, based on CODEC, to assess feature importance in a regression setting named FOCI: \textit{Feature Ordering by Conditional Independence}, which ranks a vector of covariates, $\mathbf{X}$, according to their marginal contribution in explaining a response variable $Y$ from the most to the least significant contribution. The algorithm is based on a greedy routine that, first, selects the covariate with the highest $\xi$ estimate with the response. Then, the remaining variables are evaluated based on their contribution conditioned on the first variable chosen. The routine continues until all the variables have been ranked or if the estimate provides a negative value.

\citet{chatterjee2024survey} highlights as advantages of CODEC (in random-sampling contexts): (a) its computational cost, which is of the order of $\mathcal{O}(n \log n)$, and (b) its flexibility as it does not depend on any assumptions about the underlying distribution of the variables involved, i.e., it is a model-agnostic tool.

This study seeks to explore different model-agnostic alternatives in the literature and evaluate their ability to correctly identify the number of significant lags in time series. To achieve this, the following specific objectives were proposed:
\begin{itemize}
    \item To characterize different tools for identifying $p$, including FACP, conditional Kendall's tau and Spearman's rho, and an adaptation of CODEC to the time series context.
    \item To develop a simulation routine to generate realizations of various time-indexed stochastic processes (ARMA, ARIMA, NLARMA, SETAR, ARIMA-GARCH) and, based on these simulated scenarios, assess the effectiveness of each considered technique. The complete details for the experimental setting under which the different approaches were tested are presented in Section \ref{meth}.
    \item To propose a set of recommendations to guide time series modelers on how and when to apply the most suitable technique, with an emphasis on key considerations.
\end{itemize}

The rest of this manuscript is structured as follows. Section 2 presents the methodology followed in this work for the computational simulation-based experiments. Section 3 analyzes in depth the results obtained from the simulation routines,  while Section 4 discusses the application of the proposed methodology to three benchmark datasets in time series analysis. Finally, before the references considered,  the conclusions, limitations and future work derived from this research are discussed.

\section{Methodology}
\label{meth}
Under the assumption that the association coefficient CODEC can be effectively adapted for identifying the number of relevant lags in an autoregressive process in a model-free and distribution-free environment (so that significant lags can be detected without any assumptions on the distribution of the data); we divide our methodology into two main components. The first component outlines the construction of an algorithm for lag selection based on the FOCI framework, employing CODEC as the underlying dependence measure. The second component details the implementation of this algorithm through a comprehensive simulation study, applying it to a variety of linear and nonlinear time series models to assess its empirical performance.

\subsection{Proposed Algorithm for Choosing Lags}
Let be $\left\{X_t\right\}$ an univariate times-indexed stochastic process such that, for a given value of $h$ and a time instant $t$, consider the vector of lags  $\mathbf{X} := \left({X_{t-1}, \dots, X_{t-h+1}}\right)^T$. The FOCI algorithm is then applied to identify the subset $\hat{\mathbf{S}}$ of estimated significant lags according to the CODEC measure. From this set, we take the maximum value $j$, such that $X_{t-j} \in \hat{\mathbf{S}}$; as the highest significant lag identified by CODEC. This lag can then be used as an estimator, $\hat{p}$, of the autoregressive order $p$. Nevertheless, in the simulation routines, not only the largest, but also the second and third largest lags were considered and their statistical properties as estimators of $p$ were evaluated.
\subsection{Simulation Framework}
To evaluate the performance of the proposed lag selection algorithm, we follow a structured simulation approach.  First, we considered the models presented in Table~\ref{Models}, that span a diverse set of time series behaviors, including linear and nonlinear structures combined with some or all of the following variations: (i) seasonal and trend components, (ii) moving average terms, and (iii) conditional heteroskedasticity. These include models such as SARIMA, ARIMA-GARCH, NLARMA, SETAR, and simpler processes like AR(8), providing a broad evaluation across various time series scenarios. For each model, we simulated 200 realizations for each of the following the sample sizes: $n = 100, 500, 1000, 2000$, and $5000$. To define the range of candidate lags in our lag selection procedure we use the Schwert's rule \citep{Schwert01041989}, which is a standard rule of thumb commonly used in time series analysis. The maximum number of possible lags $h^*$ is given by $$h^*=\left\lfloor{12\bigg(\cfrac{n}{100}\bigg)^{(1/4)}}\right\rfloor.$$
Within the set of lags $\left\{1,2,3,\dots,h^*\right\}$, we employed the FOCI algorithm in conjunction with the CODEC measure to determine the significant lags of each simulated time series.

\begin{table}[ht]
\centering
\resizebox{16cm}{!}{
\begin{tabular}{cl}
\toprule
\textbf{Abreviation} & \textbf{Expression}\\
\midrule
$SARIMA(2,1,1)\times (2,0,2)_{52}$ &  $(1 - 0.3 B - 0.1 B^2)(1 - 0.47 B^{52} - 0.16 B^{104})(1 - B)X_t = (1 + 0.68 B)(1 + 0.59 B^{52} + 0.62 B^{104})\varepsilon_t$ \\

ARIMA(3,1,1) & $(1 - B) X_t = 0.7 X_{t-1} - 0.5 X_{t-2} + - 0.3 X_{t-3}+\varepsilon_t + 0.4 \varepsilon_{t-1}$.\\

ARMA(3,1) & $X_t = 0.7 X_{t-1}-0.5 X_{t-2} +0.3 X_{t-3} \varepsilon_t - 0.4 \varepsilon_{t-1}$.
 \\
NLARMA(2,2) & $X_t = 2\cos(X_{t-1}) + 0.5\sin(X_{t-2}) + 0.4 \cdot \varepsilon_{t-1} + \frac{0.8}{1 + \exp(\varepsilon_{t-2})} + \varepsilon_t$.\\
SETAR(2,2;2;1) &$X_t =
\begin{cases}
2.9 - 0.4\, X_{t-1} - 0.1\, X_{t-2} + \varepsilon_t & \text{si } X_{t-2} \leq 2, \\
-1.5 + 0.2\, X_{t-1} + 0.3\, X_{t-2} + \varepsilon_t & \text{si } X_{t-2} > 2.
\end{cases}$\\


$ARIMA(1,1,1)-GARCH(1,1)$ & \quad 
$\begin{cases}
X_t = 0.75 X_{t-1} + \varepsilon_t + 0.5 \varepsilon_{t-1}, \\
\varepsilon_t = \sigma_t z_t, \quad z_t \sim \mathcal{N}(0,1), \\
\sigma_t^2 =  0.05 \varepsilon_{t-1}^2 + 0.9 \sigma_{t-1}^2.
\end{cases}$ \\

$NLAR(4)$ &  $X_t = 3\sin(X_{t-1}) + 2\sin\left(\frac{X_{t-2}}{3}\right) + 0.5\sin\left(\frac{X_{t-3}}{2}\right) - \frac{3}{1+\exp(X_{t-4})} + \varepsilon_t$ \\

$AR(8)$ & $X_t = 0.5 X_{t-1} -0.2 X_{t-2} +0.1X_{t-3}+0.2X_{t-4}+0.1X_{t-5}-0.75X_{t-6}+0.28X_{t-7} -0.25 X_{t-8} + \varepsilon_t$ \\

$SARI(5,1,0)\times (3,0,0)_{12} $ & $(1 + 0.47 B^{12} + 0.16 B^{24} - 0.74 B^{36})(1 + 0.3 B^{1}+ 0.1 B^{2} + 0.6 B^{3} - 0.2 B^{4} - 0.4 B^{5}) X_t = \varepsilon_t$ \\

$ARI(6,1,0)$ & $(1 - B) X_t =0.7 X_{t-1} -0.5X_{t-2}+0.3X_{t-3}-0.6X_{t-4}-0.25X_{t-5} -0.4 X_{t-6} + \varepsilon_t$ \\

\bottomrule
\end{tabular}
}
\caption{Selected autoregressive models for simulation.}
\label{Models}
\end{table}

As mentioned earlier, many models presented in Table~\ref{Models} contain seasonal or non-stationary components. To analyze the impact of these characteristics, the FOCI algorithm was applied not only to the raw time series but also to the pre-processed version of the series. Specifically, differencing was performed to address the presence of trends, while additive decomposition was used to remove seasonal components. This dual approach allowed for a comprehensive evaluation of the algorithm’s robustness under various data transformations, particularly in the presence of non-stationarity and marked seasonality. For each realization and each dependence measure, we considered three estimators of the autoregressive order $p$, denoted as $\hat{p}_1, \hat{p}_2$, and $\hat{p}_3$. These correspond to the three largest significant lags identified in each case. 

Finally, to analyze and evaluate the performance of the CODEC-based method, we compared it against classical approaches commonly used in time series analysis. In this case, the Pearson Correlation Coefficient and the Spearman Rank Correlation Coefficient were employed to compute the partial autocorrelation for each model across all realizations. These traditional measures provide baseline estimates of the autoregressive order $p$, serving as benchmarks for evaluating the effectiveness of the CODEC-based lag selection algorithm. The criterion used to compare the performance of the three considered auto-association measures was the Root Mean Square Error (RMSE) between the estimated and real autoregressive orders of each model.

\section{Analysis of Simulated Results}
For each one of the estimators $\hat{p}_1,  \hat{p}_2$ and $\hat{p}_3$, we obtained 200 estimates of the parameter $p$ for each combination between the sample size and autoassociation measure. When analyzing the distribution of these estimators, the results align with our expectations based on our working hypothesis, i.e, the CODEC-FOCI algorithm demonstrates empirical convergence to the true value of the  parameter in most cases. For simplicity, we first analyze the models that capture the most representative patterns observed in the simulations. Subsequently, we evaluate all simulated scenarios using the RMSE values reported in Table \ref{RMSE}.

\begin{figure}[!ht]
 \begin{subfigure}{0.495\textwidth}
    \centering
    \includegraphics[width=0.975\linewidth]{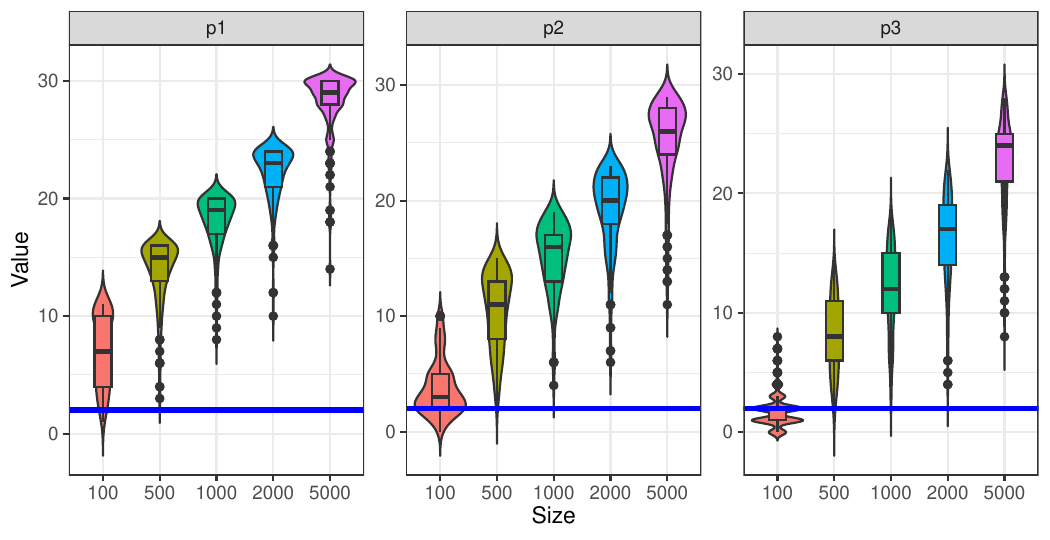}
   \caption{Pearson Correlation Coefficient.}
    \label{Serie1Pearson}
\end{subfigure}
\begin{subfigure}{0.495\textwidth}
\centering
    \includegraphics[width=0.975\linewidth]{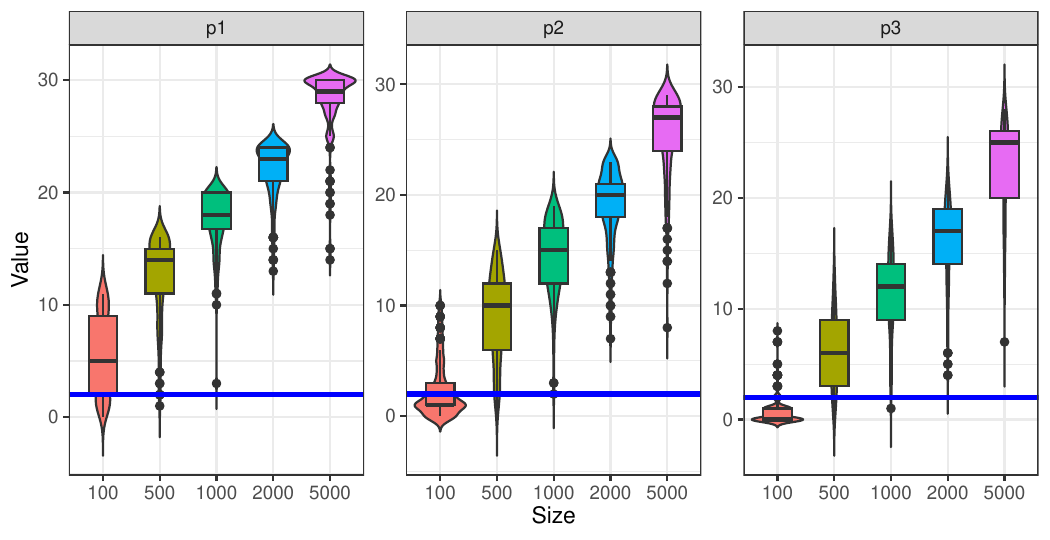}
     \caption{Spearman Rank Coefficient}
    \label{Serie1Spearman}
\end{subfigure}\\
\centering
\begin{subfigure}{0.495\textwidth}
    \centering
    \includegraphics[width=0.975\linewidth]{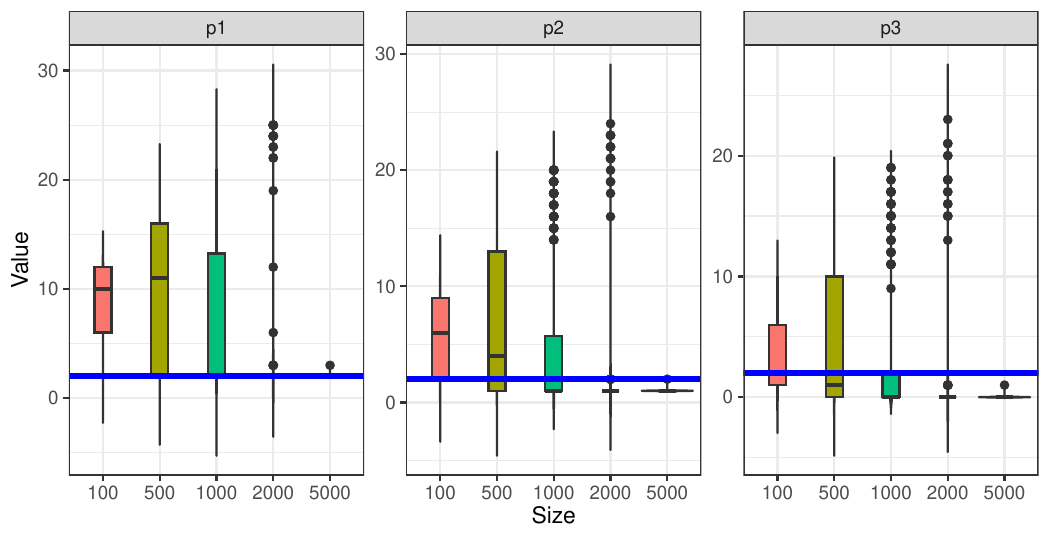}
    \caption{CODEC Coefficient with FOCI Algorithm.}
\end{subfigure}
    \caption{Distribution of realizations obtained for $SARIMA(2,1,1) \times (2,0,2)_{52}$ with respect to sample size. The true value of the parameter $p=2$ is highlighted.}
     \label{Serie1}
\end{figure}

Figure \ref{Serie1} presents a comparison of the distributions of the estimators $\hat{p}_1$, $\hat{p}_2$, and $\hat{p}_3$ across different sample sizes and autoassociation measures for the first model, $SARIMA(2,1,2)\times (2,0,2)_{52}$. In this case, the model was fitted directly to the raw data without pre-processing (e.g., detrending or seasonal adjustment). From the figure, we observe that neither the Pearson nor the Spearman measures serve as effective estimators of $p$ in this scenario. This is likely due to their high sensitivity to trend, seasonal, and moving average components. In particular, by the invertibility property, a moving average (MA) process can be expressed as an $AR(\infty)$ process \citep{box2015time}, which may distort the estimation of the autoregressive lag structure when using Pearson or Spearman-based approaches.

On the other hand, the CODEC measure shows clear convergence to the true value of the autoregressive nonseasonal parameter as the sample size $n$ increases. As expected from the construction of the estimators, in this case, $\hat{p}_1$ appears to converge to $p$, and  $\hat{p}_2$ and $\hat{p}_3$ tend to a lower value, suggesting that $\hat{p}_1$ seems to be a consistent estimator of $p$.

\begin{figure}[!ht]
 \begin{subfigure}{0.495\textwidth}
    \centering
    \includegraphics[width=0.975\linewidth]{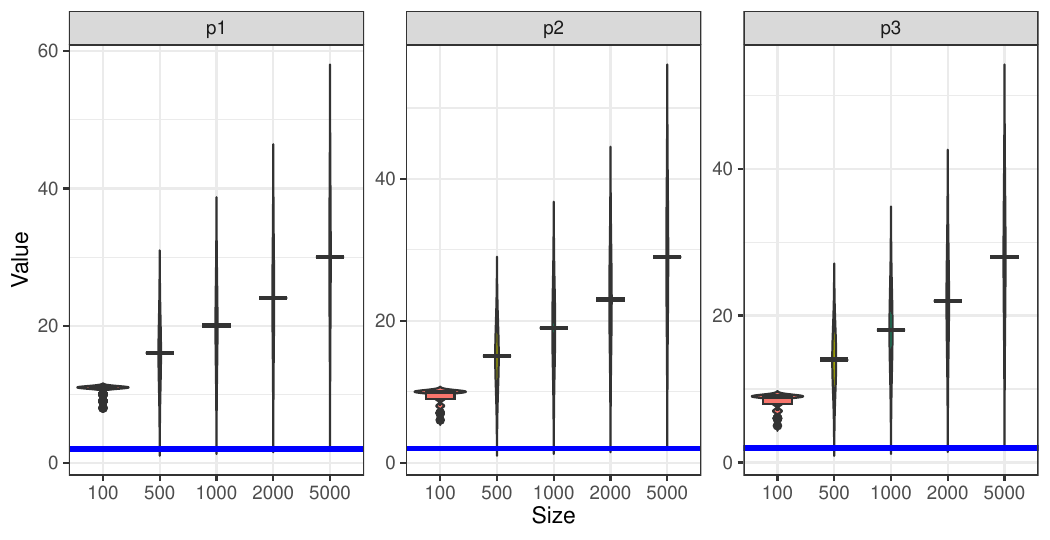}
   \caption{Pearson Correlation Coefficient.}
    \label{Serie2Pearson}
\end{subfigure}
\begin{subfigure}{0.495\textwidth}
\centering
    \includegraphics[width=0.975\linewidth]{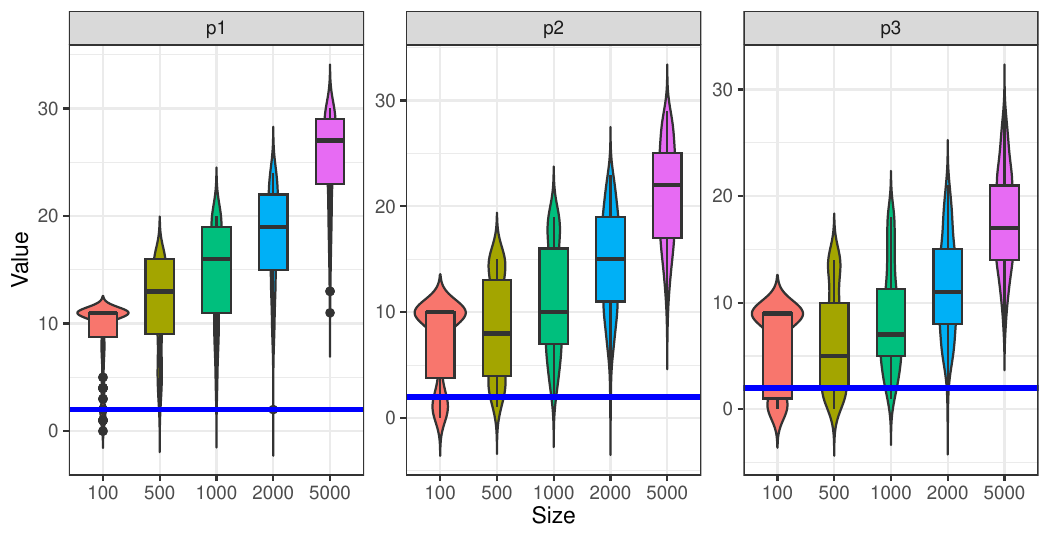}
     \caption{Spearman Rank Coefficient}
    \label{Serie2Spearman}
\end{subfigure}\\
\centering
\begin{subfigure}{0.495\textwidth}
    \centering
    \includegraphics[width=0.975\linewidth]{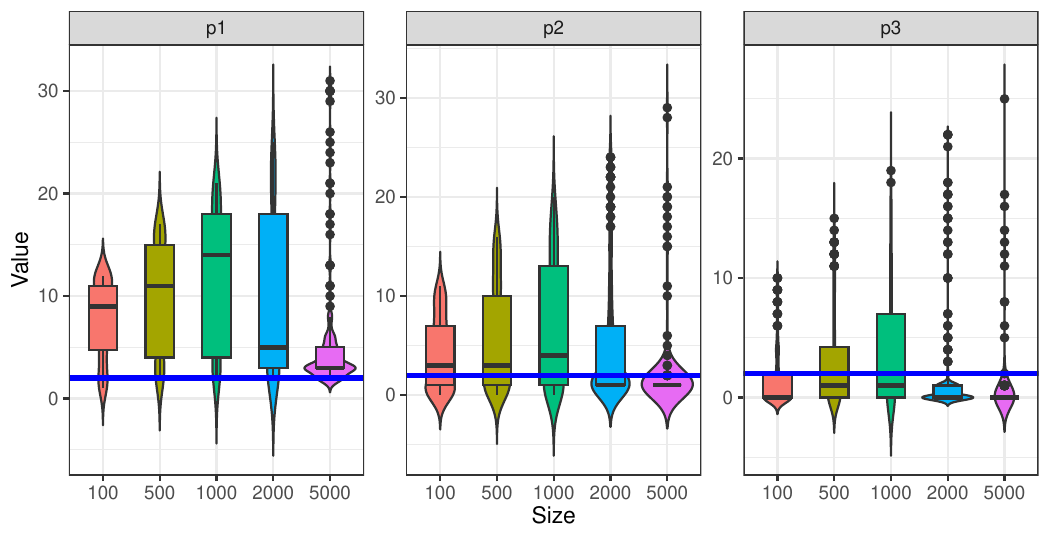}
    \caption{CODEC Coefficient with FOCI Algorithm.}
    \label{Serie2CODEC}
\end{subfigure}
    \caption{Distribution of realizations obtained for $SARIMA(2,1,1)\times (2,0,2)_{52}$ (after additive decomposition) with respect to sample size. The true value of the parameter $p=2$ is highlighted.}
     \label{Serie2}
\end{figure}

After Considering the use of classical additive decomposition as a pre-processing step for this model, Figure \ref{Serie2} shows that the distribution of the estimators tends to exhibit a reduction in variability. In this case, both Pearson and Spearman tend to estimate higher values of $p$, which may correspond to the maximum significant lag induced by the seasonal component, or potentially approaching infinity due to the invertibility property. Although the CODEC measure exhibits slightly higher variance in the distributions of $\hat{p}_1$, $\hat{p}_2$, and $\hat{p}_3$ compared to the raw data case, it still shows empirical evidence of converge towards the true value of $p$.

\begin{figure}[!ht]
 \begin{subfigure}{0.495\textwidth}
    \centering
    \includegraphics[width=0.975\linewidth]{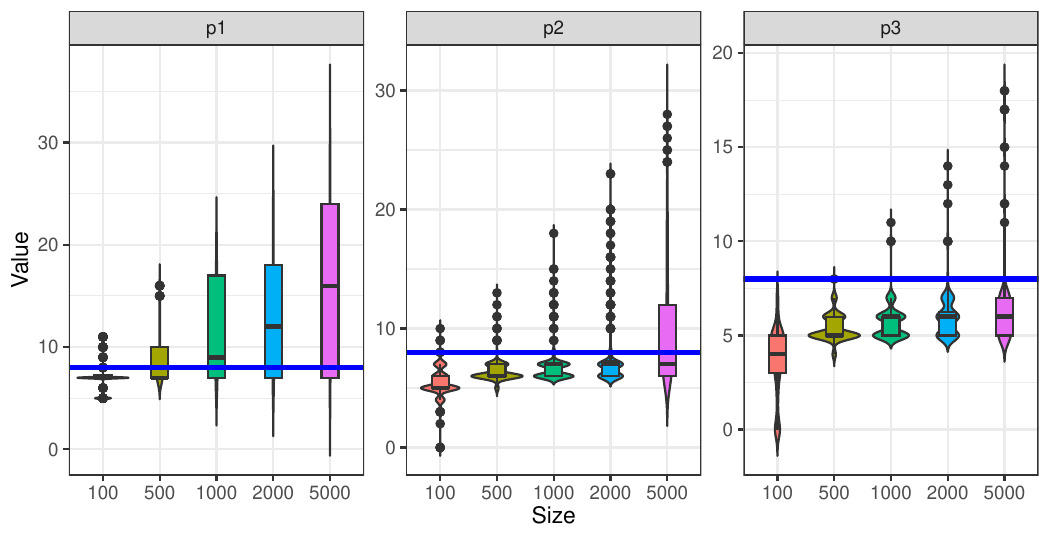}
   \caption{Pearson Correlation Coefficient.}
    \label{Serie10Pearson}
\end{subfigure}
\begin{subfigure}{0.495\textwidth}
\centering
    \includegraphics[width=0.975\linewidth]{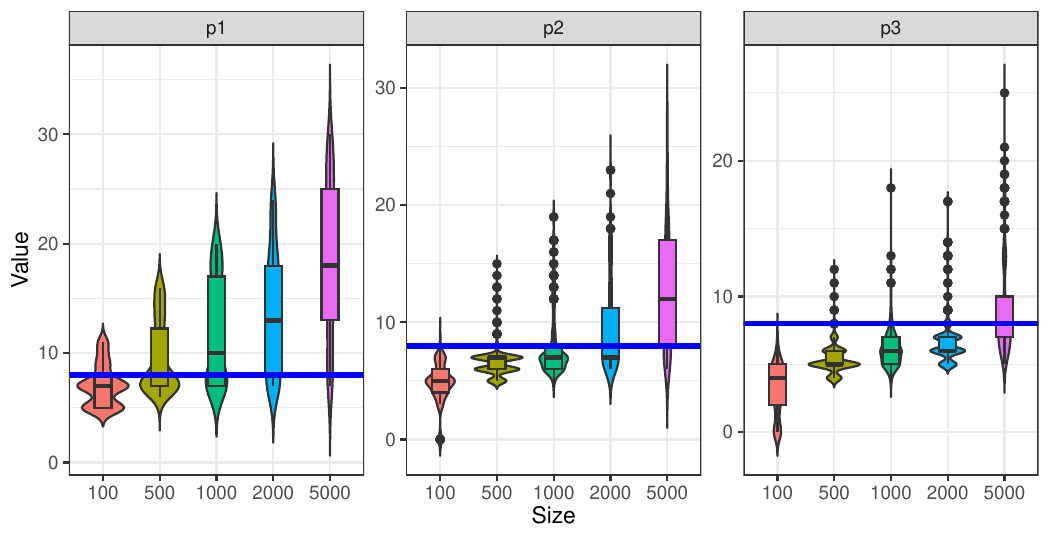}
     \caption{Spearman Rank Coefficient}
    \label{Serie10Spearman}
\end{subfigure}\\
\centering
\begin{subfigure}{0.495\textwidth}
    \centering
    \includegraphics[width=0.975\linewidth]{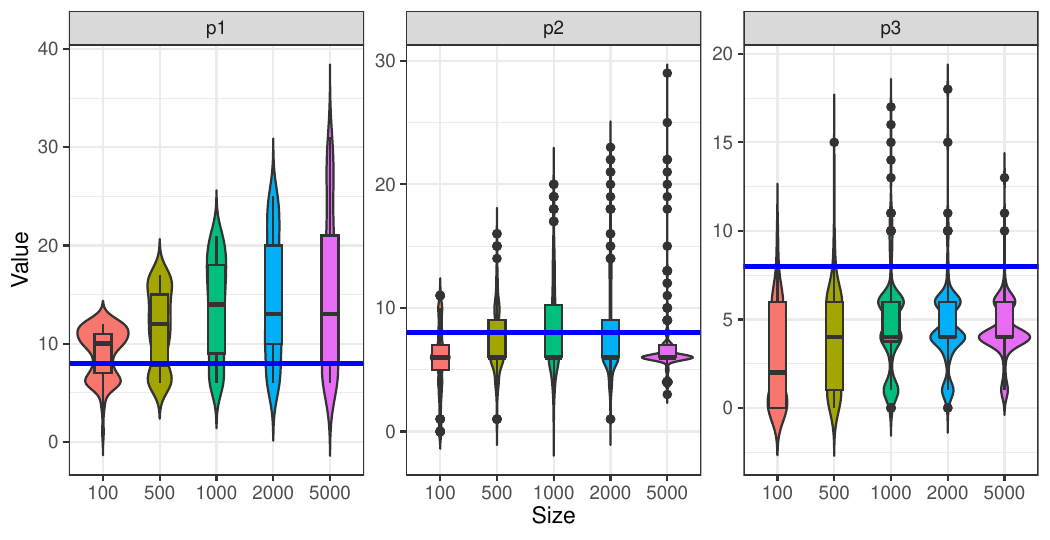}
    \caption{CODEC Coefficient with FOCI Algorithm.}
    \label{Serie10CODEC}
\end{subfigure}
    \caption{Distribution of realizations obtained for $AR(8)$ with respect to sample size. The true value of the parameter $p=8$ is highlighted.}
     \label{Serie10}
\end{figure}

Now, removing the effects of the moving average components, we considered the $AR(8)$ model, a simple and classical structure used in time series analysis. In this scenario, as shown in Figure \ref{Serie10}, the best-performing estimator was the Pearson correlation coefficient, which is consistent with expectations given the linear nature of the model. 

Nevertheless, $\hat{p}_1$ did not exhibit convergence to the true value of the parameter. This behavior could be due to the construction of $\hat{p}_1$, which selects the maximum significant lag and is therefore more sensitive to outliers, potentially explaining the lack of convergence in this case. Fortunately, the estimator $\hat{p}_2$ across all measures showed consistent convergence to $p - 1 = 7$, suggesting that $\hat{p}_2$ serves as a potentially consistent estimator for $p-1$ in the autoregressive structure, particularly under moderate noise or the presence of outliers.

\begin{figure}[!ht]
 \begin{subfigure}{0.495\textwidth}
    \centering
    \includegraphics[width=0.975\linewidth]{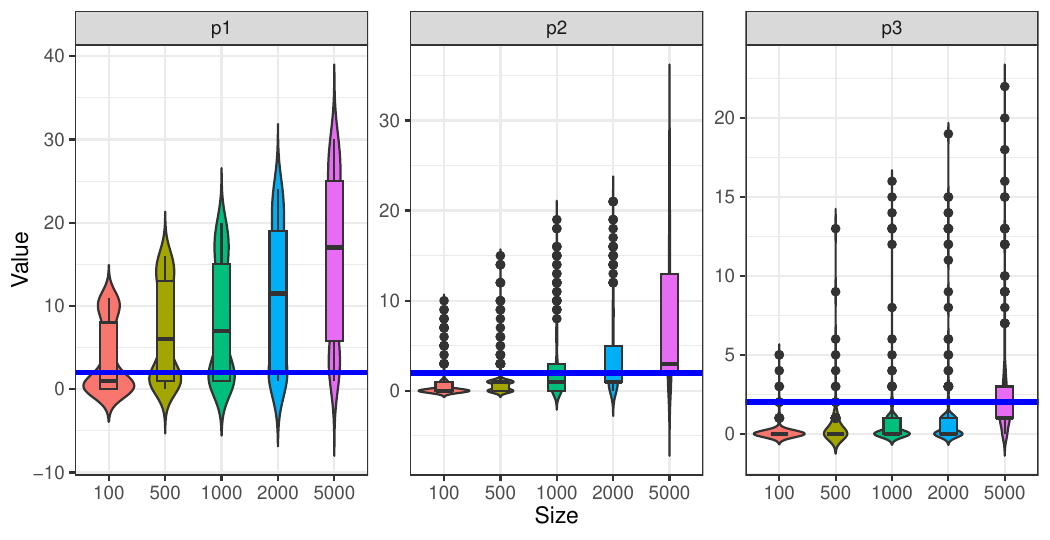}
   \caption{Pearson Correlation Coefficient.}
    \label{Serie7Pearson}
\end{subfigure}
\begin{subfigure}{0.495\textwidth}
\centering
    \includegraphics[width=0.975\linewidth]{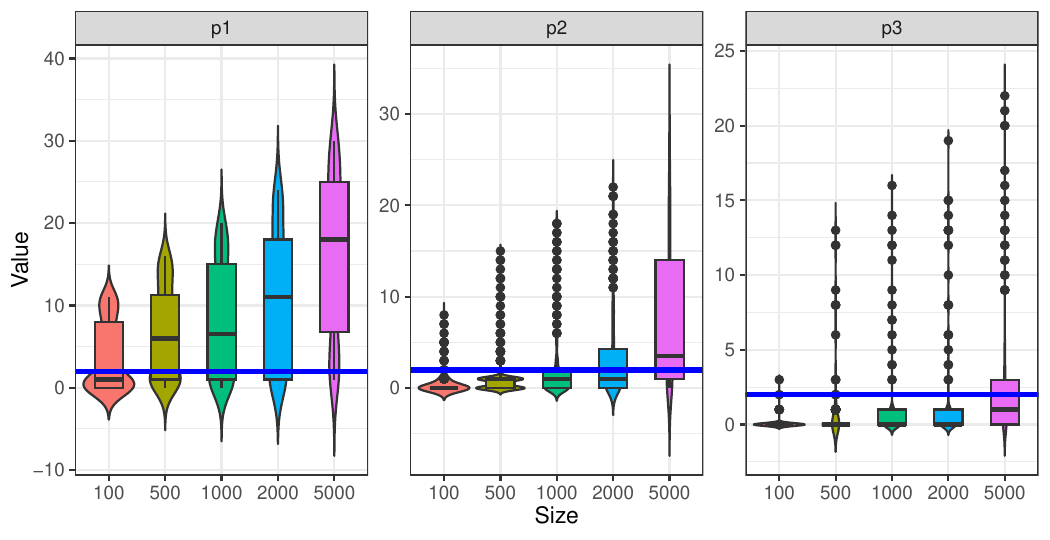}
     \caption{Spearman Rank Coefficient}
    \label{Serie7Spearman}
\end{subfigure}\\
\centering
\begin{subfigure}{0.495\textwidth}
    \centering
    \includegraphics[width=0.975\linewidth]{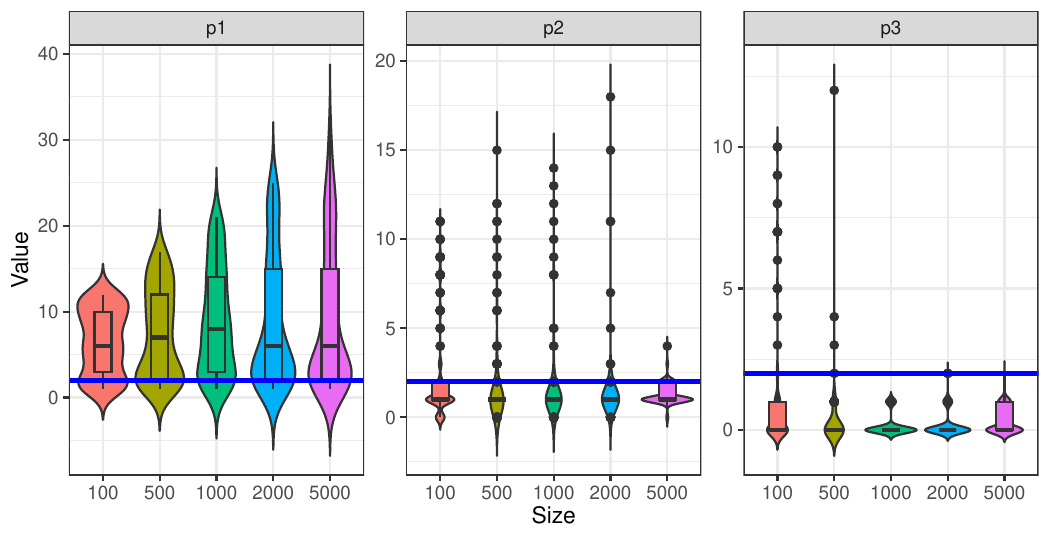}
    \caption{CODEC Coefficient with FOCI Algorithm.}
    \label{Serie7CODEC}
\end{subfigure}
    \caption{Distribution of realizations obtained for $SETAR(2,2,2;1)$ with respect to sample size. The true value of the parameter $p=2$ is highlighted.}
     \label{Serie7}
\end{figure}

We next considered the Self-Exciting Threshold Auto-Regressive Process (SETAR), a non-linear time series model. For this model, the CODEC measure demonstrated a slow convergence to the true value of $p$ (see Figure \ref{Serie7}), while the Pearson and Spearman measures did not converge to any specific value, with their variance increasing as the sample size grew. In contrast, $\hat{p}_2$, in conjunction with CODEC, exhibited a faster convergence to $p - 1 = 1$, highlighting the robustness of $\hat{p}_2$ and the CODEC-FOCI algorithm in detecting significant lags in non-linear scenarios.

Finally, the results for all the model combinations considered and different methods were aggregated and presented in Table \ref{RMSE}. The RMSE values were used to evaluate the performance and precision of each estimator based on the underlying model and simulation scenario. In the table headers, $\rho$ represents the methodology to determine $p$ based on Pearson's conditional correlation, $SP$ denotes its counterpart based on Spearman's conditional sssociation and $T_n$ denotes the CODEC method. According to the table, the Pearson and Spearman measures exhibited better performance in detecting significant lags in most scenarios with smaller sample sizes. For larger sample sizes, the CODEC estimator outperformed the others in most cases, with the exception of the $AR(8)$ model, as previously discussed.

When comparing performance across linear and non-linear models, we observe that while Pearson and Spearman performed better in linear settings with smaller sample sizes, CODEC demonstrated outstanding performance, even with limited data, when detecting significant lags in the non-linear models $NLAR(4)$ and $NLARMA(2,2)$. For larger sample sizes, CODEC successfully captured the underlying non-linear structure of the $SETAR(2,2,2;1)$ model, providing accurate lag estimations. Furthermore, the CODEC method was able to converge and identify significant lags in non-stationary processes such as $ARIMA(3,1,1)$ and $ARI(6,1)$, as well as in seasonal models like $SARIMA(2,1,1) \times (2,0,2)_{52}$ and $SARI(5,1,0)\times (3,0,0)_{12}$, even without any pre-processing.

\begin{landscape}
  \begin{table}[!ht]
        \centering
          \caption{Root Mean Square Error (RMSE) for each time series process, sample size, estimator, and autoassociation measure. The minimum (green) and second-lowest (yellow) errors obtained in each case are highlighted.}
         \vspace{1cm}
        \resizebox{22 cm}{!}{
        \begin{tabular}{cc ccc ccc ccc ccc ccc ccc ccc}
        \toprule 
        & & \multicolumn{21}{c}{\textbf{Model}} \\
        \cmidrule(rl){3-23} 
         &  & \multicolumn{3}{c}{\textbf{SARIMA}$(2,1,1) \times(2,0,2)_{52}$} &
        \multicolumn{3}{c}{\textbf{SARIMA$(2,1,1)(2,0,2)_{52}$ Decomposed}}& \multicolumn{3}{c}{\textbf{ARIMA(3,1,1) Differenced}} & \multicolumn{3}{c}{\textbf{\textbf{ARIMA(3,1,1)}}} & \multicolumn{3}{c}{\textbf{ARMA(3,1)}} &  \multicolumn{3}{c}{\textbf{NLARMA(2,2)}} & \multicolumn{3}{c}{\textbf{SETAR(2,2;2;1)}}\\
        \cmidrule(rl){3-5} \cmidrule(rl){6-8} \cmidrule(rl){9-11} \cmidrule(rl){12-14} \cmidrule(rl){15-17} \cmidrule(rl){18-20} \cmidrule(rl){21-23} 
             \textbf{Size} & \textbf{Lag} & $\rho$ & $ SP$ & $T_n$ & $\rho$ & $ SP$ & $T_n$ & $\rho$ & $ SP$ & $T_n$  & $\rho$ & $ SP$ & $T_n$ & $\rho$ & $ SP$ & $T_n$& 
             $\rho$ & $ SP$ & $T_n$ &
             $\rho$ & $ SP$ & $T_n$
             \\\midrule

             \multirow{3}{*}{100}& 1st max $p_1$&  \cellcolor{yellow} 5.89 & \cellcolor{green} 5.13 &   7.52 &
              8.76 & \cellcolor{yellow} 7.87 &  \cellcolor{green} 6.89 & 
             \cellcolor{yellow} 3.59 & \cellcolor{green} 3.38 & 6.12 &
             \cellcolor{green} 3.16 & \cellcolor{yellow} 3.70& 6.87 &
             \cellcolor{green} 3.11 & \cellcolor{yellow}  3.44 &6.45 & 
            3.97 & \cellcolor{yellow} 3.92 & \cellcolor{green} 1.57 &
            \cellcolor{yellow} 4.57 & \cellcolor{green} 4.46 & 5.87\\
             
             & 2nd max $p_2$&  \cellcolor{yellow} 2.89 & \cellcolor{green} 2.52 &  5.12 &
             7.61 & \cellcolor{yellow} 6.67&   \cellcolor{green} 4.26 &
             \cellcolor{green} 1.73 & \cellcolor{yellow} 1.89   & 3.45 &
             \cellcolor{green} 1.12 &\cellcolor{yellow}  1.96 & 4.58 & 
             \cellcolor{green} 1.62 & \cellcolor{yellow} 1.88 & 3.71 &
             2.13 & \cellcolor{yellow} 1.91 &  \cellcolor{green} 0.97 &
             \cellcolor{yellow} 2.35 & \cellcolor{green} 2.03 & 2.83\\
             
             & 3rd max $p_3$& \cellcolor{green} 1.57 & \cellcolor{yellow} 1.81 &  3.43  & 
              6.56 & \cellcolor{yellow} 5.74 & \cellcolor{green} 2.55 & 
             \cellcolor{green}2.30 & \cellcolor{yellow} 2.57 & 2.65 & 
             \cellcolor{green} 1.51 & \cellcolor{yellow}2.41 &  3.29 &
             \cellcolor{green} 2.27 & \cellcolor{yellow} 2.53 & 2.61 &
              1.88 & \cellcolor{yellow} 1.81 &  \cellcolor{green} 1.80 &
             \cellcolor{green} 1.95 & \cellcolor{yellow} 1.96 & 2.31\\ \midrule

             \multirow{3}{*}{500}& 1st max $p_1$& 12.09 & \cellcolor{yellow} 11.15 & \cellcolor{green} 9.91  &
              14.00 & \cellcolor{yellow}10.57 & \cellcolor{green} 9.66  &
             \cellcolor{yellow} 6.00 & \cellcolor{green}5.97 &  7.83 &
             \cellcolor{green} 6.09 &\cellcolor{yellow}  6.81 & 8.86 &
             \cellcolor{green} 6.35 & \cellcolor{yellow} 6.37 &  8.63 &
             6.95 & \cellcolor{yellow} 6.42 & \cellcolor{green} 1.93 & 
             \cellcolor{yellow} 7.43 & \cellcolor{green}7.17  & 7.56\\
             
             & 2nd max $p_2$&  9.17 & \cellcolor{yellow} 8.08 & \cellcolor{green} 7.69 & 
             13.00 & \cellcolor{yellow} 8.05 & \cellcolor{green} 6.32 &
             \cellcolor{green} 2.23 & \cellcolor{yellow} 2.60  & 3.60 &
             \cellcolor{green} 2.16 & \cellcolor{yellow} 3.31 & 7.07 &
             \cellcolor{green} 2.34 & \cellcolor{yellow} 3.05 & 3.89 &
            2.97 & \cellcolor{yellow} 2.85 & \cellcolor{green} 0.78 &
             3.23 & \cellcolor{yellow} 3.11 & \cellcolor{green} 2.34\\
             
             & 3rd max $p_3$& 6.76 & \cellcolor{green} 5.37& \cellcolor{yellow}5.84&
             12.00 &\cellcolor{yellow}6.31&  \cellcolor{green} 4.25 &
            \cellcolor{green}1.43 & \cellcolor{yellow} 2.29 &2.42 &
            \cellcolor{green}1.09 & \cellcolor{yellow} 1.93 & 5.52 &
            \cellcolor{green} 1.48 & 2.26 & \cellcolor{yellow} 2.24 & 
            \cellcolor{yellow}1.63 & \cellcolor{green} 1.50 & 1.64 &
            \cellcolor{yellow} 2.07  & 2.41 &  \cellcolor{green} 2.00  \\ \midrule
             
          \multirow{3}{*}{1000}& 1st max $p_1$& 16.26 &  \cellcolor{yellow} 15.83 & \cellcolor{green} 9.04  &
           18.00 & \cellcolor{yellow} 13.67 & \cellcolor{green} 11.92 &
           \cellcolor{green} 8.77 & \cellcolor{yellow}9.14 & 10.13 &
           \cellcolor{yellow}9.15 & 10.78 & \cellcolor{green} 6.09 &
           \cellcolor{yellow} 8.77 & \cellcolor{green} 8.76 & 10.10 &
           9.97 & \cellcolor{yellow} 9.18 & \cellcolor{green} 1.83 &
           9.35 & \cellcolor{yellow} 9.20 & \cellcolor{green} 9.19  \\

             & 2nd max $p_2$& 13.52 &  \cellcolor{yellow}12.93  & \cellcolor{green} 7.67 &
             17.00 &  \cellcolor{yellow} 10.28 & \cellcolor{green} 8.22 &
             \cellcolor{yellow}3.79 & 4.07   & \cellcolor{green} 3.66 &
             \cellcolor{green} 4.07 & 5.53 & \cellcolor{yellow} 5.44 &
            4.25 & \cellcolor{yellow} 3.84 & \cellcolor{green} 3.78 & 
            \cellcolor{yellow} 4.02 & 4.03 & \cellcolor{green} 0.76 &
            4.66 & \cellcolor{yellow} 4.53 & 
            \cellcolor{green} 2.13 \\

             & 3rd max $p_3$& 10.54 &  \cellcolor{yellow} 10.09  & \cellcolor{green} 6.42 &
             16.00 & \cellcolor{yellow} 7.97 &  \cellcolor{green} 5.60 & 
             \cellcolor{green} 1.77 &  2.00 &  \cellcolor{yellow} 1.90 & 
            \cellcolor{green} 1.66  &  \cellcolor{yellow} 2.23 & 5.10 &
            \cellcolor{green} 1.64 & 2.32  &  \cellcolor{yellow} 1.85 &
             \cellcolor{yellow} 1.87 & 1.88 & \cellcolor{green} 1.58 &  
             2.84 &  \cellcolor{yellow} 2.77 &  \cellcolor{green} 1.88 \\ \midrule
              \multirow{3}{*}{2000}& 1st max $p_1$& \cellcolor{yellow} 20.28 &  \cellcolor{yellow} 20.17 & \cellcolor{green} 6.20 &
              22.00 & \cellcolor{yellow} 16.85 & \cellcolor{green} 11.04 &
              \cellcolor{green} 11.64 &  11.83 & \cellcolor{yellow}11.80
              & \cellcolor{yellow} 11.92 & 14.06& \cellcolor{green} 0.93 &
              \cellcolor{green}  12.06&  \cellcolor{yellow} 12.17 & 12.38 &
              \cellcolor{yellow} 12.54 & \cellcolor{yellow} 12.52 &\cellcolor{green}  1.35 &
              12.32 & \cellcolor{yellow} 12.15& \cellcolor{green} 10.29\\

             & 2nd max $p_2$ &  17.52 & \cellcolor{yellow} 17.30 & \cellcolor{green} 5.47 &
             21.00 & \cellcolor{yellow} 13.53 & \cellcolor{green} 9.38 &
             5.68 &  \cellcolor{yellow} 4.97 & \cellcolor{green} 2.07 & 
              \cellcolor{yellow} 6.24 &  7.94 &  \cellcolor{green} 1.89 & \cellcolor{yellow} 5.65 &  5.74 &  \cellcolor{green} 3.76 &
             6.17 & \cellcolor{yellow} 6.13 &  \cellcolor{green} 0.67  &
             5.85 & \cellcolor{yellow} 5.49&
               \cellcolor{green} 2.02 \\

             & 3rd max $p_3$ & 14.74 & \cellcolor{yellow} 14.55 & \cellcolor{green} 4.88 &
             20.00 & \cellcolor{yellow} 10.48 & \cellcolor{green} 5.02  &
             \cellcolor{yellow} 2.06 & 2.65 &  \cellcolor{green} 1.28 &
             \cellcolor{yellow} 2.89  & 4.13 & \cellcolor{green}  2.88
             & 2.90 & \cellcolor{yellow} 2.58 & \cellcolor{green} 1.71&
             2.75 & \cellcolor{yellow} 2.70 & \cellcolor{green} 1.41 &
             3.40 & \cellcolor{yellow} 3.30 & \cellcolor{green} 1.86
             \\ \midrule

              \multirow{3}{*}{5000}& 1st max $p_1$& \cellcolor{yellow}  26.25 &  26.32 &  \cellcolor{green} 0.07 &
              28.00 & \cellcolor{yellow} 23.67 & \cellcolor{green} 7.41 &
              15.40 & \cellcolor{yellow} 15.28 &  \cellcolor{green} 13.40 &
              18.21 & \cellcolor{green} 7.64 & \cellcolor{yellow} 11.21 &
              \cellcolor{yellow} 15.34 & 19.60 &  \cellcolor{green} 0.83 &
              16.82 & \cellcolor{yellow} 15.70 & \cellcolor{green} 13.66 &
              17.49 & \cellcolor{yellow} 16.91 & \cellcolor{green} 1.44
              \\

             & 2nd max $p_2$ & \cellcolor{yellow} 23.56 &  23.69 &  \cellcolor{green} 0.99 &
             27.00 &  \cellcolor{yellow} 19.61  & \cellcolor{green} 4.66 &
             8.29 & \cellcolor{yellow} 8.02 & \cellcolor{green} 1.47 &
             \cellcolor{yellow} 8.67 & 13.40 &  \cellcolor{green} 1.75 & 
             9.12 & \cellcolor{yellow} 8.17 &  \cellcolor{green} 2.26 &
            10.12 &  \cellcolor{yellow} 9.16 & \cellcolor{green} 0.57 &
            9.59 &  \cellcolor{yellow} 9.58 &   \cellcolor{green} 0.92
             \\
             
            & 3rd max $p_3$ & \cellcolor{yellow} 21.00 & 21.40 &  \cellcolor{green} 1.99 &
             26.00&  \cellcolor{yellow} 16.47 & \cellcolor{green} 3.29 &
             \cellcolor{yellow} 3.53 &  4.21 & \cellcolor{green} 1.36 & \cellcolor{yellow} 4.32 & 8.43 &  \cellcolor{green} 2.73 &
             4.74& \cellcolor{yellow} 4.26 & \cellcolor{green} 1.35 &
             4.98 & \cellcolor{yellow}  4.66 &\cellcolor{green}  1.24 &
             \cellcolor{yellow} 4.04&  4.34&\cellcolor{green}  1.79 
             \\\bottomrule
             
             \toprule 
             
        & & \multicolumn{21}{c}{\textbf{Model}} \\
        \cmidrule(rl){3-23} 
         &  & \multicolumn{3}{c}{\textbf{ARIMA(1,1,1)-GARCH(1,1)}} &
        \multicolumn{3}{c}{\textbf{NLAR(4)}}& \multicolumn{3}{c}{\textbf{AR(8)}} & \multicolumn{3}{c}{\textbf{SARI}$ (5,1,0) \times (3,0,0)_{12}$} & \multicolumn{3}{c}{\textbf{SARI}$ (5,1,0) \times (3,0,0)_{12}$\textbf{ decomposed}} &  \multicolumn{3}{c}{\textbf{ARI(6,1,0) Differenced}} & \multicolumn{3}{c}{\textbf{ARI(6,1,0)}}\\
        \cmidrule(rl){3-5} \cmidrule(rl){6-8} \cmidrule(rl){9-11} \cmidrule(rl){12-14} \cmidrule(rl){15-17} \cmidrule(rl){18-20} \cmidrule(rl){21-23} 
             \textbf{Size} & \textbf{Lag} & $\rho$ & $ SP$ & $T_n$ & $\rho$ & $ SP$  & $T_n$ & $\rho$ & $ SP$ & $T_n$  & $\rho$ & $ SP$ & $T_n$ & $\rho$ & $ SP$ & $T_n$& 
             $\rho$ & $ SP$ &  $T_n$ &
             $\rho$ & $ SP$ & $T_n$
             \\\midrule

             \multirow{3}{*}{100}& 1st max $p_1$& \cellcolor{yellow} 4.77 & \cellcolor{green} 4.70 & 7.22 &
              \cellcolor{yellow} 3.60 &  3.64 &\cellcolor{green} 3.37 & 
             \cellcolor{green} 1.79 &  \cellcolor{yellow} 2.07 & 2.73 &
             \cellcolor{yellow}4.34& \cellcolor{green} 3.54 &  5.52 &
             4.89  &  \cellcolor{green}3.66 &  \cellcolor{yellow}4.73 &
             \cellcolor{green} 2.10 & \cellcolor{yellow} 2.22 &3.87
             & \cellcolor{green} 1.86 &  \cellcolor{yellow} 2.40 & 3.98\\
             
             & 2nd max $p_2$&  \cellcolor{yellow} 1.91 & \cellcolor{green}1.81 & 3.74 &
              \cellcolor{yellow} 3.29 &  3.43 &  \cellcolor{green} 3.12 & 
             \cellcolor{green} 3.05 &3.50 &  \cellcolor{yellow}3.31  &
              \cellcolor{yellow} 2.65 &\cellcolor{green} 2.30 & 3.65 &
             \cellcolor{green} 3.05& 4.06& \cellcolor{yellow} 3.67 &
              \cellcolor{yellow}2.55 &  \cellcolor{green} 2.49 & 3.29 &
             \cellcolor{green} 1.97 &\cellcolor{yellow}  2.60 & 4.06  \\
             
             & 3rd max $p_3$& \cellcolor{green} 0.91 & \cellcolor{yellow} 0.94 & 2.32 &
             \cellcolor{green} 3.74 & \cellcolor{yellow} 3.81  &  3.90 &
             \cellcolor{green} 4.68 & \cellcolor{yellow} 5.20 & 5.84 &
             \cellcolor{green} 1.96 & \cellcolor{yellow} 2.77 & 3.49 & 
             \cellcolor{green} 1.95 & 4.63 & \cellcolor{yellow}  4.04 &
             \cellcolor{yellow} 4.13 & \cellcolor{green} 4.03 & 4.17 &
             \cellcolor{green} 3.95 & \cellcolor{yellow} 4.23 & 4.87\\ \midrule

             \multirow{3}{*}{500}& 1st max $p_1$& \cellcolor{green} \cellcolor{yellow} 7.95 & \cellcolor{green} 7.43 & 9.14 &
             5.70 & \cellcolor{yellow} 5.64 & \cellcolor{green} 3.50 &
             \cellcolor{green} 2.90& \cellcolor{yellow} 3.64 &5.23  &
              10.95 &\cellcolor{yellow} 10.71 &  \cellcolor{green} 9.25 &
             10.53 & \cellcolor{yellow} 6.20 &  \cellcolor{green} 3.04 &
             \cellcolor{green}3.26 &\cellcolor{yellow}  4.48 &6.32 &
              \cellcolor{green} 4.26 & 5.16 &\cellcolor{yellow} 4.30\\
             
             & 2nd max $p_2$& 3.84 & \cellcolor{green} 3.07 & \cellcolor{yellow} 3.20 &
             3.23 & \cellcolor{yellow} 3.13 &  \cellcolor{green}2.84 &
             \cellcolor{green} 1.87 &\cellcolor{yellow} 2.05 &2.93  &
             9.68 & \cellcolor{yellow} 8.86 & \cellcolor{green} 6.76  &
             9.29  &\cellcolor{green} 4.02 & \cellcolor{yellow} 4.08 &
             \cellcolor{yellow} 2.53 & \cellcolor{green} 2.25 & 3.64 &
             \cellcolor{green} 1.89 & \cellcolor{yellow} 2.18 & 4.16\\
             
             & 3rd max $p_3$& 1.69 & \cellcolor{green} 1.38& \cellcolor{yellow}1.40&
             \cellcolor{green} 3.41 & \cellcolor{yellow} 3.55 & 3.79 &
             \cellcolor{green}2.72&\cellcolor{yellow} 2.84 & 4.95 &
             7.61 & \cellcolor{yellow}6.87 & \cellcolor{green} 5.08 &
             7.87  & \cellcolor{green} 3.66 & \cellcolor{yellow} 5.03 &
            \cellcolor{yellow}3.36 & \cellcolor{green} 2.80 & 3.77 &
            \cellcolor{yellow} 2.69 &  \cellcolor{green} 1.57 &   5.04\\ \midrule
             
          \multirow{3}{*}{1000}& 1st max $p_1$&  \cellcolor{yellow} 10.63 & \cellcolor{green} 10.34 &  11.84 &
          7.87 &   \cellcolor{yellow} 7.73 &\cellcolor{green}2.74 &
            \cellcolor{green} 6.06 & \cellcolor{yellow} 6.36 &7.43  &
           14.58 &\cellcolor{yellow}13.73 & \cellcolor{green} 11.44 &
            14.77 & \cellcolor{yellow} 9.31 &  \cellcolor{green} 2.92 &
           \cellcolor{green}6.32 &  \cellcolor{yellow} 6.95 &8.47  &
            \cellcolor{yellow} 6.32 & 7.52 &  \cellcolor{green} 3.43\\

             & 2nd max $p_2$& 5.44 &  \cellcolor{yellow}4.91  & \cellcolor{green} 4.27 &
             \cellcolor{yellow}3.59 &  3.66 & \cellcolor{green} 2.48 &
             \cellcolor{yellow} 2.11 & \cellcolor{green} 2.90   & 4.05 &
             12.94  &  \cellcolor{yellow}12.06 & \cellcolor{green} 9.23 &
             13.52 &\cellcolor{yellow} 5.60 & \cellcolor{green} 4.00 & 
            \cellcolor{green} 2.99 & \cellcolor{yellow} 3.02 & 3.93 &
           \cellcolor{green} 2.60 & \cellcolor{yellow} 3.80 &  4.13\\

             & 3rd max $p_3$& 2.74 & \cellcolor{yellow} 2.31  & \cellcolor{green} 1.82 &
            \cellcolor{green}  2.71 & \cellcolor{yellow} 2.77 & 3.48 &
             \cellcolor{yellow} 2.46 & \cellcolor{green} 2.40 & 4.39 &
             11.68 &\cellcolor{yellow} 10.80 & \cellcolor{green} 7.31 & 
             12.30 & \cellcolor{green} 3.87 & \cellcolor{yellow} 5.00 &
            \cellcolor{yellow} 3.17 & \cellcolor{green} 2.43  & 3.40 & 
           \cellcolor{yellow} 2.33 & \cellcolor{green} 2.10 & 5.05 \\ \midrule

              \multirow{3}{*}{2000}& 1st max $p_1$&  13.90 & \cellcolor{green} 13.46 &  \cellcolor{yellow} 13.89 &
              11.18 & \cellcolor{yellow} 11.16 & \cellcolor{green} 0.63 &
             \cellcolor{green} 7.89 & \cellcolor{yellow} 8.04 &  9.05 &
             18.99 &  \cellcolor{yellow} 18.91 & \cellcolor{green} 11.73 &
             18.92 & \cellcolor{yellow} 15.84 &  \cellcolor{green} 3.00 &
              \cellcolor{green} 8.64 &\cellcolor{yellow} 9.20 & 10.78 &
              \cellcolor{yellow} 8.88 &  11.50 & \cellcolor{green} 2.65
              \\

             & 2nd max $p_2$ & 7.38 &  \cellcolor{yellow} 6.79 &  \cellcolor{green} 4.06 & 
             5.75 & \cellcolor{yellow} 5.57 &  \cellcolor{green} 2.10 & 
             \cellcolor{green} 3.67 & \cellcolor{yellow} 4.07 &  4.09 & 
             17.22 &  \cellcolor{yellow} 16.88 &  \cellcolor{green} 9.71 & 
            17.74 &   \cellcolor{yellow} 12.05 & \cellcolor{green} 4.00 &  
             \cellcolor{green} 4.42 &  \cellcolor{yellow} 4.47 & 4.84 & 
             \cellcolor{green} 3.88 & 6.98 & \cellcolor{yellow} 4.01 \\ 
             
             & 3rd max $p_3$ & 3.94 & \cellcolor{yellow} 3.46 &  \cellcolor{green} 1.13 & 
             \cellcolor{green} 3.03 & 3.34 & \cellcolor{yellow} 3.11 & \cellcolor{green} 2.35 &\cellcolor{yellow} 2.51 & 4.17 & 
             15.90& \cellcolor{yellow} 15.34 & \cellcolor{green} 8.08 &
             16.54 &\cellcolor{yellow}  8.72 &\cellcolor{green}  5.00 &
             3.33&  \cellcolor{green} 2.18&  \cellcolor{yellow} 3.22 &
             \cellcolor{green} 2.41 & \cellcolor{yellow} 3.76 & 5.02
             \\ \midrule
             
             \multirow{3}{*}{5000}& 1st max $p_1$& 19.15 & \cellcolor{yellow} 19.49 &\cellcolor{green} 16.84 &
             14.51 &  \cellcolor{yellow} 14.30 &  \cellcolor{green} 0.31 &
             \cellcolor{yellow} 11.94 & 12.72 & \cellcolor{green} 10.67 &
             25.00 &\cellcolor{yellow} 24.99 & \cellcolor{green} 7.11&
              \cellcolor{yellow} 24.84 &  25.00 &
             \cellcolor{green} 15.11 &
             13.85 & \cellcolor{yellow} 13.76 & \cellcolor{green} 13.61 &
             \cellcolor{yellow} 14.05 & 16.90 & \cellcolor{green} 2.19\\

             & 2nd max $p_2$  &\cellcolor{yellow} 10.99&  11.19 &  \cellcolor{green} 3.33 &
              \cellcolor{yellow} 7.25 & 7.56 &   \cellcolor{green} 1.97 &
             \cellcolor{yellow} 5.57 &6.91&  \cellcolor{green} 3.72 & 
             24.00 & \cellcolor{yellow} 23.98 &  \cellcolor{green} 6.28 &
              \cellcolor{yellow} 23.77&24.00 &  \cellcolor{green} 4.00 &
             \cellcolor{yellow} 7.52 &   7.72 &  \cellcolor{green} 4.95 &
              \cellcolor{yellow} 7.31 &  10.25 &  \cellcolor{green} 3.99\\

             & 3rd max $p_3$ &  5.94 & \cellcolor{yellow}  5.16 &  \cellcolor{green} 0.43&
             3.65 &  \cellcolor{yellow} 3.50 &  \cellcolor{green} 2.96 & 
            \cellcolor{green} 2.82 &  \cellcolor{yellow} 3.88 & 3.95 & 
            \cellcolor{yellow} 22.41 &  22.68 & \cellcolor{green} 5.88 & 
              \cellcolor{yellow} 22.71 &22.99&  \cellcolor{green} 5.00 & 
             \cellcolor{yellow}  3.45 &  3.73 &  \cellcolor{green} 2.63 &
              \cellcolor{green} 3.12& 6.31 &\cellcolor{yellow} 4.99
             \\\bottomrule
        \end{tabular}
        }
        \label{RMSE}
    \end{table}
    \end{landscape}

Comparing the estimators $\hat{p}_1$, $\hat{p}_2$, and $\hat{p}_3$, we observe a consistent pattern in their performance when used in conjunction with the CODEC measure: the more non-linear the model, the better the performance of $\hat{p}_1$. Naturally, this relationship is influenced by the sample size, with $\hat{p}_1$ converging to the true value of $p$ as $n$ increases. In contrast, $\hat{p}_3$ performed better in linear models, while $\hat{p}_2$ proved particularly effective for the $SETAR(2,2,2;1)$ model and consistently showed the second-best performance across most scenarios. Overall, $\hat{p}_1$ delivered outstanding results in both seasonal and non-linear contexts. This performance hierarchy may be explained by the relative robustness of $\hat{p}_2$ and $\hat{p}_3$ compared to $\hat{p}_1$, especially in the presence of noise or outliers.

\begin{figure}[!ht]
    \centering
    \includegraphics[width=0.85\linewidth]{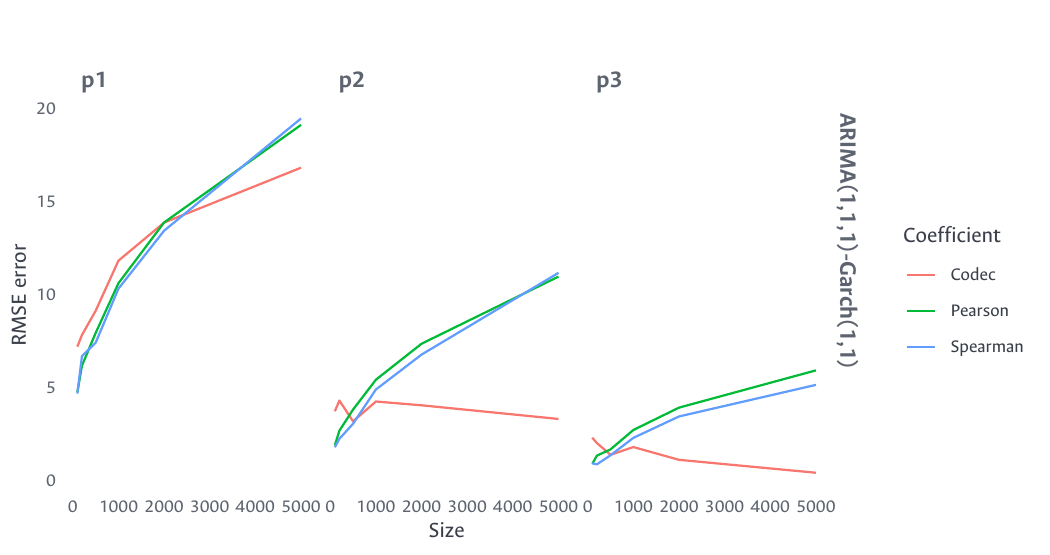}
    \caption{Errors obtained for each autoassociation measure in the $ARIMA(1,1,1)-GARCH(1,1)$ model.}
    \label{ErrorsGarch}
\end{figure}

The ARIMA(1,1,1)–GARCH(1,1) model had a behavior distinct from all other models considered. The heteroskedastic nature of the process may have adversely impacted the performance of the autoassociation measures, leading to a considerable number of apparent ``outliers." These, in turn, increased the variance and interfered with the convergence properties of the estimator $\hat{p_1}$. As previously noted, the robustness of $\hat{p}_2$ and $\hat{p}_3$ when paired with the CODEC-FOCI algorithm allowed them to extract meaningful lag information despite the non-constant variance. Their corresponding RMSE values support this convergence behavior, as illustrated in Figure~\ref{ErrorsGarch}.

\begin{figure}[!ht]
 \begin{subfigure}{0.495\textwidth}
    \centering
    \includegraphics[width=0.975\linewidth]{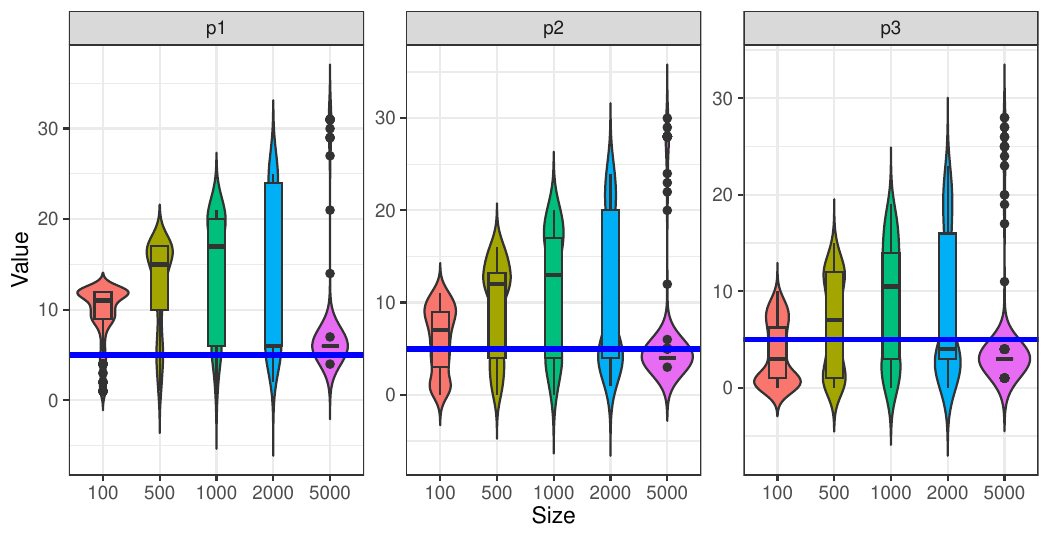}
   \caption{Raw-data.}
    \label{Serie11}
\end{subfigure}
\begin{subfigure}{0.495\textwidth}
\centering
    \includegraphics[width=0.975\linewidth]{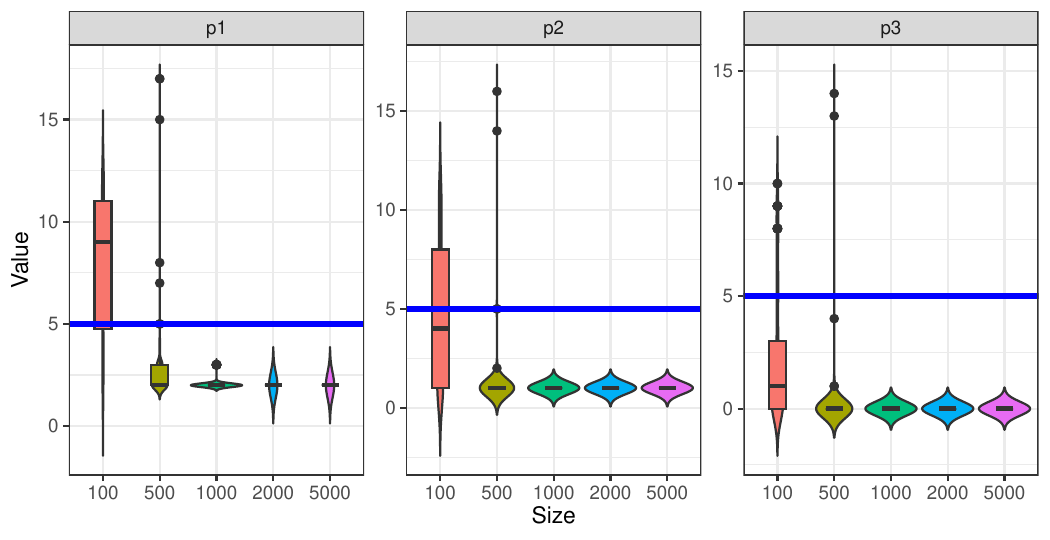}
     \caption{Decomposed model}
    \label{Serie12}
\end{subfigure}
    \caption{Comparison between the distributions of realizations obtained for $SARIMA(5,1,0)\times (3,0,0)_{12}$ and CODEC measure with and without additive  decomposition. The true value of the parameter $p=5$ is highlighted.}
     \label{Serie1112Comparison}
\end{figure}

\begin{figure}[!ht]
    \centering
    \includegraphics[width=0.75\linewidth]{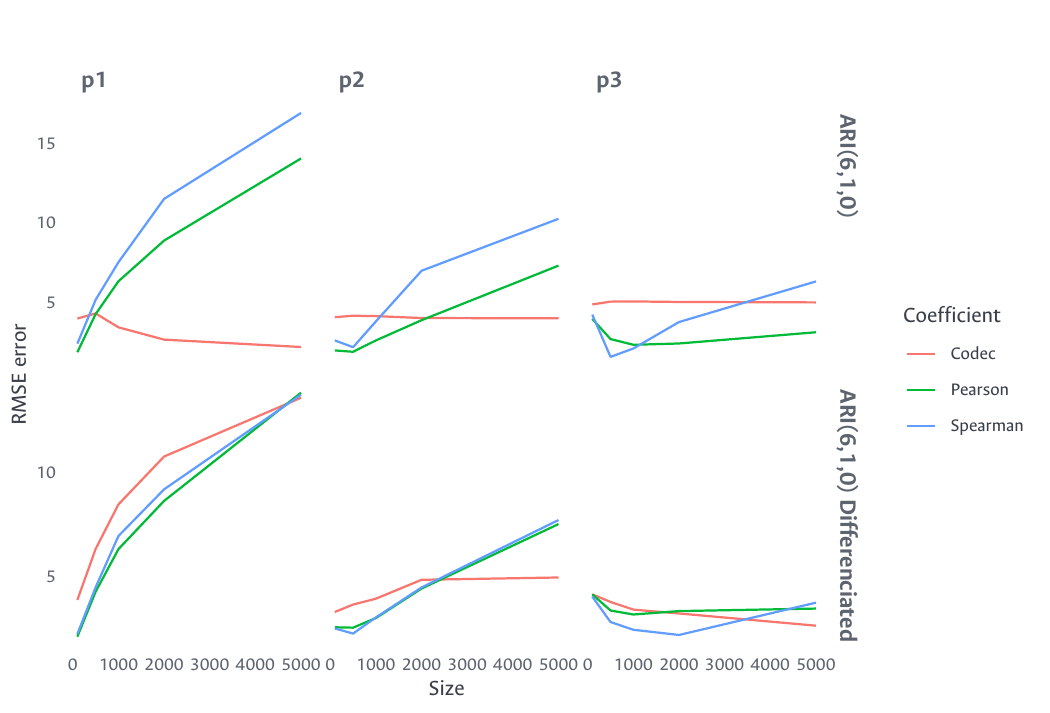}
    \caption{Comparison between the errors obtained for each autoassociation measure in the $ARI(6,1)$ model with differenced and non-differenced data.}
    \label{ErrorsARI}
\end{figure}

Regarding trend and seasonality pre-processing, Table~\ref{RMSE} indicates that classical decomposition adversely impacts the performance of the lag detection estimators when used in conjunction with the CODEC measure. Specifically, it increases the variance in the distribution of the estimated lag values. This effect is evident in Figure~\ref{Serie2} for the $SARIMA(2,1,1) \times (2,0,2)_{52}$ model and in Figure~\ref{Serie1112Comparison} for the $SARI(5,1,0)\times (3,0,0)_{12}$  model, where the estimators show convergence toward the seasonal autoregressive components $P = 2$ and $P = 3$, respectively. However, this apparent convergence may be caused by the non-parametric decomposition process rather than a consequence of the estimator or the algorithm itself.
In the same manner, differencing the data to remove trend components appears to negatively affect the performance of the CODEC-FOCI method. Figure~\ref{ErrorsARI} compares the RMSE behavior of the estimators for the $ARI(6,1)$ model when applied to differenced and non-differenced data. Based on the results in Table~\ref{RMSE}, it is evident that convergence does not occur, or requires significantly larger sample sizes, when working with differenced data.

\section{Application to Benchmark Datasets}

Three widely used datasets were employed to contrast the results provided by the methodology proposed in this work with the different models fitted to these series. The first series is the count of annual sunspots \citep[p. 471]{tong1990non}, the second series is the number of lynx specimens captured in Canada \citep[p. 557]{brockwell1991time} and the third one is the monitoring of the amount of international airline passengers \citep[p. 669]{box2015time}. Some basic features of these series are presented in Table \ref{tab:classic-ts-models}.

\begin{longtable}{@{}p{6.5cm}p{9.5cm}@{}}
\label{tab:classic-ts-models} \\
\toprule
\textbf{Series} & \textbf{Models Fitted} \\
\midrule
\endfirsthead

\toprule
\textbf{Series} & \textbf{Models Fitted} \\
\midrule
\endhead

\textbf{Annual Sunspot Numbers} 
\begin{itemize}
  \item \textbf{Frequency:} Annual
  \item \textbf{Time Period:} 1700-1987
  \item  \textbf{Sample size:} 288 observations
  \item \textbf{Location:} Global (Zürich observatory)
\end{itemize} 
& 
\begin{enumerate}
  \item AR(3) - \citet[p. 264]{box2015time}
  \item AR(2), AR(9), SETAR(2, 11, [8]) - \citet[pp. 420 - 428]{tong1990non}
  \item FAR(8,3)  - \citet[p. 329]{fan2008nonlinear}
  \item EXPAR(9)  - \citet{azouagh2019exponential}
\end{enumerate} \\
\midrule

\textbf{Annual Number of Lynx Trapped} 
\begin{itemize}
  \item \textbf{Frequency:} Annual
  \item \textbf{Time Period:} 1821-1934
    \item  \textbf{Sample size:} 114 observations
  \item \textbf{Location:} Canada (Hudson’s Bay Company)
\end{itemize} 
& 
\begin{enumerate}
  \item AR(4) - \citet[p. 426]{box2015time}
  \item SETAR(2,2,[2]) - \citet[p. 102]{tong1983threshold}
  \item FAR(2,2), SETAR(2,2,[2]), AR(2) - \citet[p. 328]{fan2008nonlinear}
  \item FFNN(1,2;2) - \citet{kajitani2005forecasting}
\end{enumerate} \\
\midrule

\textbf{Monthly Airline Passengers} 
\begin{itemize}
  \item \textbf{Frequency:} Monthly
  \item \textbf{Time Period:} 1949-1960
    \item  \textbf{Sample size:} 144 observations
  \item \textbf{Location:} USA (International airline passengers)
\end{itemize} 
& 
\begin{enumerate}
  \item SARIMA(0,1,1)$\times$(0,1,1)$_{12}$ - \citet[p. 359]{box2015time}
  \item LSTM (1 seasonal difference, 12-period input) - \citet[p. 317]{brownlee2018deep}
\end{enumerate} \\
\bottomrule
\caption{Summary of classic time series data, their characteristics, and fitted models in the literature} 
\end{longtable}

Table \ref{tab:classic-ts-models} also includes some models fitted to the data in seminal textbooks and articles to give an idea of the structures, and more importantly, the configuration of lagged values considered to predict the given series at a time point of interest. It is worth mentioning that not all authors followed the same procedure or used the same portions of data to elicit the model structures shown; therefore, these models are considered for reference only. To clarify the notation used, AR($p$) denotes an autoregressive model of $p$ lags; SETAR($k$,$p$,[$r$]) denotes a self-exciting threshold autoregressive model with $k$ regimes, at most $p$ lags in each regime, and as threshold variable the observation $r$ units in the past; FAR($p$,$d$) is a functional-coefficient autoregressive model with $p$ lags and each coefficient is a function of at most $d$ lags; EXPAR($p$) is an exponential autoregressive model with $p$ lags; SARIMA($p$,$d$,$q$)$\times$($P$,$D$,$Q$)$_{12}$ is a monthly seasonal autoregressive integrated moving average model whose regular component is model with $p$ lags, $d$ differences and $q$ lags of the noise processes. Analogously, its seasonal component is composed of $P$ seasonal lags, $D$ differences and $Q$ seasonal lags of the noise processes. Finally, a FFNN($l$,$n$;$p$) model stands for a feedforward neural network with $l$ hidden layers, $n$ hidden nodes per layer and $p$ lags used as input; and LSTM model stands for a long short-term memory neural network.

\begin{figure}[!ht]
	\centering
       \includegraphics[width=0.7\linewidth]{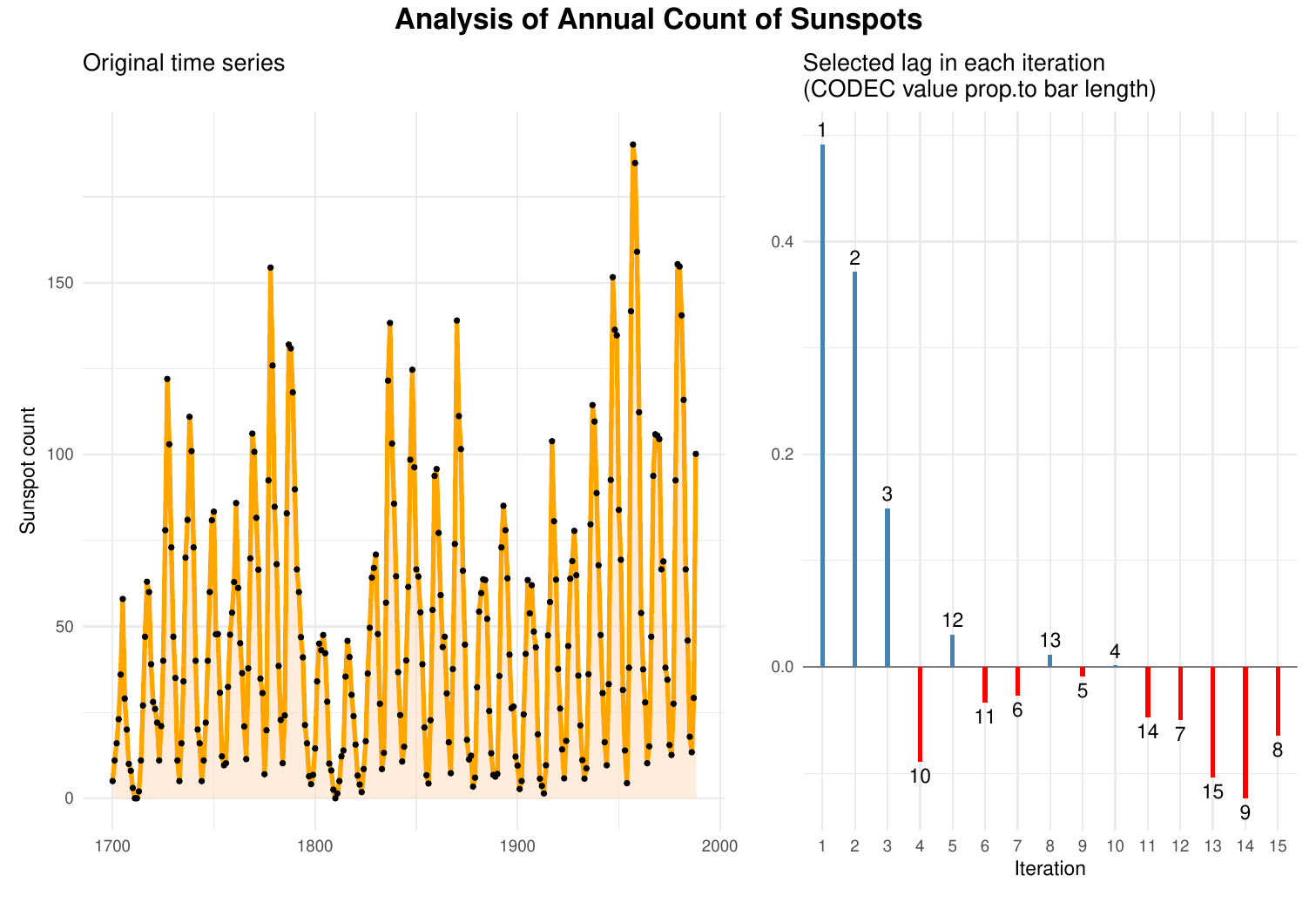}
	\caption{Results of the proposed methodology when applied to the Sunspot time series.}
	\label{fig_spot}
\end{figure}

\begin{figure}[!ht]
	\centering
       \includegraphics[width=0.7\linewidth]{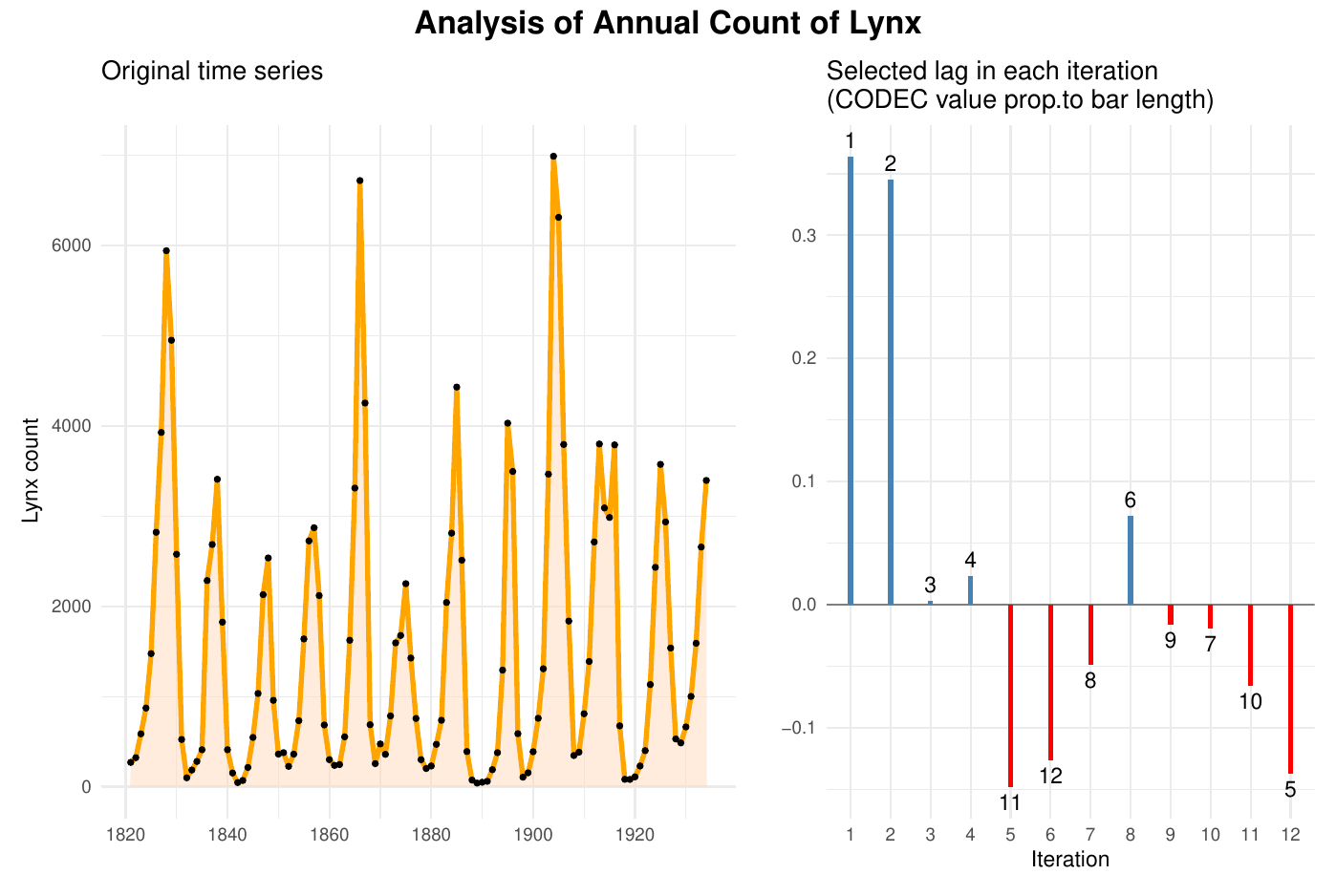}
	\caption{Results of the proposed methodology when applied to the Lynx time series.}
	\label{fig_lynx}
\end{figure}

\begin{figure}[!ht]
	\centering
       \includegraphics[width=0.7\linewidth]{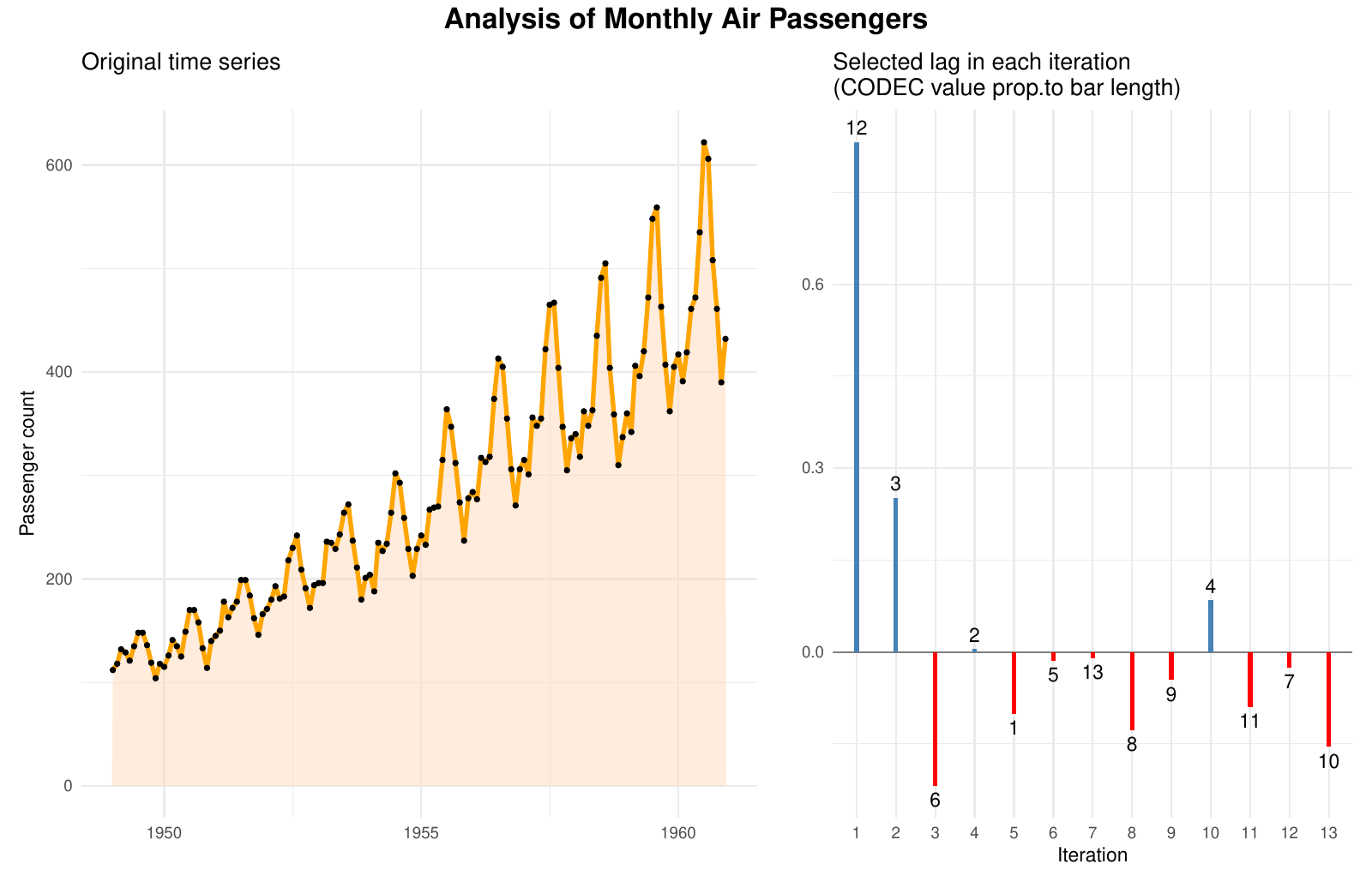}
	\caption{Results of the proposed methodology when applied to the Passengers time series.}
	\label{fig_pssg}
\end{figure}

The results from the FOCI-CODEC implementation on the selected time series are consistent with the structures given in the textbooks and articles shown in Table~\ref{tab:classic-ts-models}. Figures~\ref{fig_spot},~\ref{fig_lynx},~\ref{fig_pssg} show the selected lag at each iteration of the algorithm for these time series. It is worth reminding that FOCI is supposed to stop once it finds a negative estimate of the conditional CODEC coefficient. However, the re-order of all lags considered is presented for completeness

In the sunspots series, the FOCI algorithm stops at the third iteration, selecting the first three lags as the most significant to construct the autoregressive model, i.e., $\hat{p}_{\text{CODEC}} = 3$. In the case of the Lynx series, the algorithm stops at the fourth iteration, selecting the first four lags to model $X_t$, i.e., $\hat{p}_{\text{CODEC}} = 4$. Finally, for the passengers time series, the algorithm stops at the second iteration, selecting the third and twelfth lags; hence, $\hat{p}_{\text{CODEC}} = 12$, detecting the seasonal component.

\section{Conclusions, Limitations and Future Work}
In this article, we implemented the recently proposed conditional dependence measure by \citet{azadkia2021simple}, known as CODEC, in the context of time series, as a model-free alternative to identify the number of lags to include in an autoregressive model. The implementation was carried out by adapting the FOCI algorithm proposed by the authors to detect the most significant lag relationships within the series. To evaluate the performance of the proposed methodology, a series of simulations were conducted, analyzing the behavior of the FOCI algorithm under different sample sizes and time series models, and contrasting its performance with classical auto-association procedures in time series. The time series models considered in the simulation framework include structures with a diverse range of behaviors, such as moving average components, seasonal and trend effects, conditional heteroskedasticity, and both linear and nonlinear dynamics.

The results of our simulation study indicate that the FOCI-based approach can effectively estimate the structural parameter $p$ of a time series model in most scenarios, particularly when the series is non-stationary or exhibits nonlinear relationships. In such settings, the CODEC measure outperformed traditional correlation-based approaches, such as Pearson and Spearman, which are commonly used in time series analysis to identify relevant lags. Moreover, in many of the models considered, the FOCI algorithm demonstrated improved detection of significant lags even without pre-processing steps such as differencing or additive decomposition. A possible explanation for this robustness lies in the rank-based nature of CODEC: by working with data ranks rather than raw values, the methodology may attenuate the influence of nonstationary effects, thereby yielding more stable and reliable lag identification across a wide variety of processes.

Nevertheless, the simulation study also exposed certain limitations of the FOCI algorithm, such as a weak performance in time series characterized by conditional heteroskedasticity, as observed in GARCH-type models. Additionally, the estimator $\hat{p}_1$ appeared to be sensitive to outliers and extreme values, which is a natural consequence of its definition. In such cases, $\hat{p}_2$ and $\hat{p}_3$ demonstrated greater robustness in estimating $p$, although they exhibited bias in most of the scenarios considered.

The algorithm was also applied to three benchmark datasets: the Annual Sunspot Numbers \citep{tong1990non}, the annual number of lynx trapped in Canada \citep{brockwell1991time} and the monthly international airline passengers \citep{box2015time}. This application aimed to assess the behavior of the methodology on well-known and widely studied time series. The estimations of the structural parameter $p$ obtained by the FOCI algorithm in each case were consistent with those reported in the reviewed seminal textbooks and articles, reinforcing the practical applicability of the proposed approach.

In conclusion, the implementation of the FOCI algorithm in time series analysis has shown notable relevance and potential applicability for detecting complex relationships between lags in autoregressive models. It is important to emphasize that this study is primarily exploratory, and some of the observed findings require further investigation to assess their validity and limitations across a broader range of scenarios.

Several directions for future research emerge from this work. For instance, a theoretical study of the convergence, statistical power, and other properties of the FOCI algorithm would be of great value. Additionally, the implementation and comparison of the FOCI algorithm with other rank-based association measures could provide a more comprehensive understanding of its performance. Another promising avenue is the potential adaptation of the algorithm to estimate the structural parameter of Moving Average models, $q$. While a thorough exploration of these topics lies beyond the scope of the present paper, they constitute relevant and interesting areas for future investigation.

\bibliography{refs}
\end{document}